# Spin Dynamics in Graphene and Graphene–like Nanocarbon Doped with Nitrogen : the ESR Analysis

*Ashwini P. Alegaonkar[1], Arvind Kumar[2], Satish K. Pardeshi[1] and Prashant S. Alegaonkar[2]*

[1]*Dept. of Chemistry, University of Pune, Pune, 411 007, MS, India*

[2]*Dept. of Applied Physics, Defence Institute of Advanced Technology, Girinagar, Pune 411 025, MS, India*

**ABSTRACT**

Nano–engineered spin degree of freedom in carbon system may offer desired exchange–coupling with optimum spin–orbit interaction which is essential, to construct solid state qubits, for fault–tolerant quantum computation. The purpose of this communication is to analyze spin dynamics of, basically, four types of systems: (i) Graphene (system with inversion symmetry), (ii) Graphene–like nanocarbons (GNCs, broken inversion symmetry and heterostructure, $sp^2+ sp^3$, environment), and (iii) their nitrogen doped derivatives. The spin transport data was obtained using the electron spin resonance spectroscopy (ESR) technique, carried out over 123–473K temperature range. Analysis of shape, linewidths of dispersion derivatives,, and g–factor anisotropy has been carried out. Spin parameters such as, spin–spin, $T_{sl}$, spin–lattice, $T_{sl}$, relaxation time, spin–flip parameter,$|b|$, spin relaxation rate, $\Gamma_{spin}$, momentum relaxation rate, $\Gamma$, pseudo chemical potential, $\tilde{\mu}$, density of states, $\rho$, effective magnetic moment, $\mu_{eff}$, spin–, defect–concentration, and Pauli susceptibility, $\chi_{spin}$ has been estimated, and examined for their temperature as well as interdependence. Details of the analysis are presented. The quantitative study underlined the following facts: (i) by and large, spin dynamics in Graphene and GNCs is significantly different, (ii) transport of spin behaves in opposite fashion, after doping nitrogen, in both the systems, (iii) reduction in the magnetization has been observed for both GNCs and Graphene, after doping nitrogen, (iv) hyperfine interactions have been observed in all classes of systems except in GNCs, (v) nitrogen doped GNCs seems to be appropriate for qubit designing.

## I. INTRODUCTION

Harnessing quantum laws in nanocarbon system seems to be promising for information processing that outperforms their classical counterparts for unconditional, secured communication.[1] Such quantum systems interact with their environment via up or down spin of electron, in the form of strings of quantum bits (qubits).[2-3] In general, the spin degree of freedom of electron endowed to its lattice is on the *talking term* and could be exploited, primarily, using spin–orbit–(SO) and exchange–interactions.[4] Nonetheless, the crucial requirements on spin density wave propagation are quantum entanglement,[5] coupling the spin to lattice vibration,[6] coherence,[7] and hyperfine interaction of the electron spin with the surrounding nuclear spin.[8] In Graphene, contributions to SO interaction is due to intrinsic,[9] Bychkov–Rashba,[10] ripple,[4] and extrinsic[11]. These interactions are supposed to be weak, in μeV regime,[12] due to low atomic number of carbon.[4] The exchange–interactions originating due to spin–flip and spin splitting are predicted to be identical for disordered– and ordered–$sp^2$ Graphene network. In fact, these interactions are supposed to independent of SO coupling.[4] The spin–lattice relaxation time is estimated in 10–100 ns range[13] with no or low hyper fine interactions, due zero–spin $^{12}C$ nuclei (abundance ≈ 99.0–98.9%).[14] In recent experiments, the spin–lattice relaxation time is measured as short as 60–150 ps.[15-16] The Klein tunnelling paradox showed *leaky* spin density wave propagation due to difficulty in creating gap–tuneable Graphene quantum dot.[17] The tunnel–couple dots has found hindrance in Heisenberg–spin exchange–coupling due to valley degeneracy that exist in the vicinity of Dirac point in Graphene.[18] The role of strength and contribution of SO components is still a subject of discussion,[4,9,11,12] whereas, significant emphasis has been given on the study of virtual spin degree of freedom,[19-22] neglecting the fourfold spin–degenerated states of the actual spin in the Graphene band structure.[23] Realistic Graphene contain inversion, broken inversion symmetry, intrinsic, and extrinsic heterostructure lattice environment. And, present scenario demands experimental validation for studying spin transport in realistic Graphene super lattice which could address a few issues mentioned above. Principally, spin relaxation time is inseparable from SO interactions and has direct dependence on transport parameters like spin relaxation rate, spin flip probability, momentum relaxation rate, relaxation time, etc. Analyzing them may develop fundamental insight in designing tuneable spin degrees of freedom; crucial for favourable SO– and exchange–interactions.

The focus of this communication is to study spin dynamics of Graphene, Graphene–like nanocarbons (GNCs), and their nitrogen doped derivatives. The spin transport data was obtained using the electron spin resonance spectroscopy (ESR) technique, carried out over a temperature range 123–473K, on the bulk powder specimen. Spin transport parameters has been estimated, examined, and analyzed for their temperature as well as interdependence. The study revealed that, there is significant difference in the spin transport properties of Graphene and GNCs, moreover, transport of spin behaves in opposite fashion after doping nitrogen. The overall reduction in the magnitude of magnetization has been observed, after nitrogen doping in both the systems. The hyperfine interactions have been observed in all classes of systems except GNCs. Understanding spin transport may provide clue for construction of solid state qubits which is crucial for future quantum computing devices. Details are presented.

## II. EXPERIMENTL DETAILS

Basically, we have two types of carbon environments (i) Graphene, pure $sp^2$ carbon (i.e. system with inversion symmetry), and (ii) GNCs, disordered $sp^2$ network (broken inversion symmetry and heterostructure). In the present work, Graphene–like nanocarbon sheets (GNCs) were synthesized according to protocol reported.[24] The material obtained by our synthesis methodology was graphene–like nanocarbon and not graphene. The basic structure of GNCs is of carbophane family in which carbon is in more than one hybridization state i.e.

mixture of $sp^2$ and $sp^3$ carbon atoms. In such structures carbocyclic molecule contain covalently bridged aromatic rings called phanes.[42] This system is typically 2–5 layers and each layer contains heavy local disorder in the form of configuration and missing carbon atoms. While synthesizing, transformation of ¼ of the $sp^2$ graphitic carbon to sp3 carbon takes place so that remaining ¾ are all in the benzoid confirmation and within $sp^2$ cluster no $sp^3$ carbon present. It has presence only at the adjacent site to $sp^2$ cluster in each layer. Every $sp^3$ is covalently bonded within $sp^3$ zone and at the interface to deviates from covalent nature while bonding with the $sp^2$ carbon, in the form of three benzoid ring. The remaining covalent bond to each $sp^3$ carbon provides covalent interlayer bond. Hence, we have observed minimum 2 layers of GNCs. Details are provided in the reference mentioned above.

*Treatment with tetrakis(dimethylamino) ethylene (TDAE) for nitrogen doping*

Nitrogen doping was carried out using tetrakis(dimethylamino) ethylene (TDAE) compound. Initially, the suspension of GNCs was prepared in 25 ml of tetra hydro furan (THF) and 0.1 mg of TDAE was added in GNCs suspension. After adding TDAE, the suspension was sonicated for 30 min followed by room temperature stirring for 8–10 h. The suspension was allowed to settle for about 5 h. The vacuum filtration was carried out using PTFE filter (pore size ~1.2 μm). The GNCs treated with TDAE was termed as N–GNCs. In similar fashion, Graphene and N–Graphene were prepared. The powder obtained was used for examined for structure–property relationship using preliminary characterization such as Fourier transform infrared–, and UV–Visible–spectroscopy. Details are provided in Supporting information.

*Electron spin resonance measurements on the systems*

The electron spin resonance (ESR) measurements were carried out using a standard ESR set up equipped with electromagnet, microwave bridge, resonant cavity, waveguide circuitry and spectrometer consol. Initially, a known amount of sample, under investigation, was placed in the rectangular cavity consisting of cylindrical sample transfer ports. The mode of the cavity was $TE_{102}$. The sample was positioned at cavity centre where the magnetic component of the microwave standing wave attains the local maxima. This configuration provides maximum sensitivity to the measurement. The microwave energy was injected and coupled via an iris screw. The critical coupling, achieved using iris screw controlled the amount of incident and reflected microwave radiation in the cavity. The measurements were performed at microwave frequency of ~ 9.1 GHz. (X–band). The maximum microwave output was varied from sample to sample in the range 985 µW to 1000 µW. Further, for receiver mode, phase shifter enables one to match the phase of microwave signal reflected from cavity with the phase of microwave injected in the reference arm of the circulator. The value of variable phase shifter was kept fixed at zero. The quality factor of the cavity was 12000. In general, the ESR signal is weak and submerge in the background level noise. By enhancing the sensitivity of spectrometer one can amplify the obtained ESR signal. In the present measurements, the modulation frequency was kept constant at 100 kHz by adjusting the band pass filter parameter of the lock–in–amplifier. The static magnetic field was swept slowly at a spectrum–point time constant 0.1 sec over the range 300 mT to 370 mT with the amplitude of modulation frequency kept at 6 kHz. The field centre was 336 mT and ESR line width, $\Delta H = 0.05$ mT. The signal–to–noise ratio was computed to be 20 at 300K with 1s per spectrum–point constant. The sensitivity of the system was $7.0 \times 10^{9}$ spins / 0.1 mT and resolution 2.35 µT. All measurements were carried out at a temperature range 123–473K. And the first derivative of the paramagnetic absorption signal was recorded for samples. The pristine Graphene and GNCs after doping nitrogen were designated, respectively, as N–Graphene and N–GNCs. The highest concentration samples were taken from each batch of N–Graphene and N–GNCs.

### III. RESULTS, ANALYSIS, AND DISCUSSION

Basically, we have two types of carbon environments (i) Graphene, pure $sp^2$ carbon (i.e. system with inversion symmetry), and (ii) GNCs, disordered $sp^2$ network (broken inversion symmetry and heterostructure). Particularly; GNCs carbon phanes contains mixture of $sp^2$ and $sp^3$ carbon. Such systems are sensitive to interband optical transitions which occur between 1–3 eV. The characteristic vibration band in Raman spectrum shows peak at 1670 cm$^{-1}$ with full width half maximum 100 cm$^{-1}$.[43] These are typically 2–5 layered system in which clusters of $sp^2$ are isolated by $sp^3$ carbon zones having random patterning in them. What nitrogen doping dose? It provides is the extrinsic environment in addition to intrinsic symmetry that exists in two distinct systems.

### A. Line width, and shape–the results

A great deal of information can be obtained from the careful analysis of the width and shape of a resonance absorption line. Two common line shapes are Gaussian and Lorentzian. FIG. 1 shows ESR dispersion derivatives, $\frac{dY}{dH}$ as a function of magnetic field. The upper panel is ESR for Graphene and N–Graphene. The spectrum is found to be asymmetrical for Graphene, over the measured temperature regime. For both, the fitted line shape is found to be intermediate between Gaussian and Lorentzian.

For Graphene, the line–width is found to be varied from 1.1027 ± 0.0071 (123K) to 1.2311 ± 0.0063 mT (473K). For N–Graphene, the degree of asymmetry of the ESR line is increased and the width in the range 1.1495 ± 0.0066 (123K) to 1.3217 ± 0.0035 mT (473K). After nitrogen doing in Graphene, there is slight increase in the line–width with inhomogeneous broadening. The inhomogeneous broadening is indicative of three sources (i) more than two spin–lattice relaxation time, $T_{sl}$, are merged into one overall line, (ii) charge inhomogeneities (the so called puddles). The charge inhomogeneities over the volume of the sample exceeds the natural line–width, $\frac{1}{\gamma T_{sl}}$, where $\gamma$ is gyromagnetic ratio and prevent relaxing electron from reaching the Dirac point, and have the average minimal charge density as $10^9$ cm$^{-2}$. The spins associated with these charge inhomogeneities, in various parts of the sample, find themselves in different field strength, and the resonance is broadened in an inhomogeneous manner.[25] The source (iii) is hyperfine interactions (HFI), the weak magnetic interactions between the spin of unpaired electrons and nuclear spin. The interaction could be its own nucleus or ensemble of nearest neighboring nuclei spins. Lower panel in FIG. 1 shows ESR spectra for GNCs and N–GNCs. For GNCs, the fitted line shape is found to be perfectly Lorentzian, whereas, for N–GNCs mix feature, Lorentzian–Gaussian. The line–widths are somewhat narrowed down after doping nitrogen to GNCs. Their magnitudes are 1.2964±0.0083 to 1.2669±0.0091, for GNCs, and 0.7091±0.0049 to 0.6710±0.0038, for N–GNCs, over the measured temperature range. The signature of hyperfine structure is observed in Graphene as well as N–Graphene including N–GNCs. These weak lines are zoomed in each inset of FIG.1. These lines are not accounted for by considering forbidden transition. In general, the emergence of HFI in graphene comes from the nuclear spins of the $^{13}$C isotope, having natural abundance ≈ 1–1.1 %. The dominant $^{12}$C has zero spin. These pairs of lines are separated from allowed lines by nuclear resonance frequency of the proton, at the field $H_r$ used for the ESR measurement. The interaction is present over entire region of magnetic field and at all temperatures. The presence of HFI indicates that such interaction could originate from presence of adatom of odd nuclei like $^{1}$H. The weak lines arise from matrix of protons which undergo spin–flip when electron spins of surrounding trapped hydrogen atoms are reoriented. The coupling is dipolar and the intensity of the weak lines varies approximately as : $H_r^{-2}$ and, there could be $(2I + 1)^n$ hyperfine components for *n* nuclei with the nuclear spin I. The total number of hyperfine component, $N_{hfs}$, could be: $N_{hfs} = \prod_i (2I_i + 1)^{n_i}$, where the symbol, $\prod_i(I)$, denotes the formation of a product.[26] Except GNCs, other system has shown additional weak lines, present over the swept magnetic field.

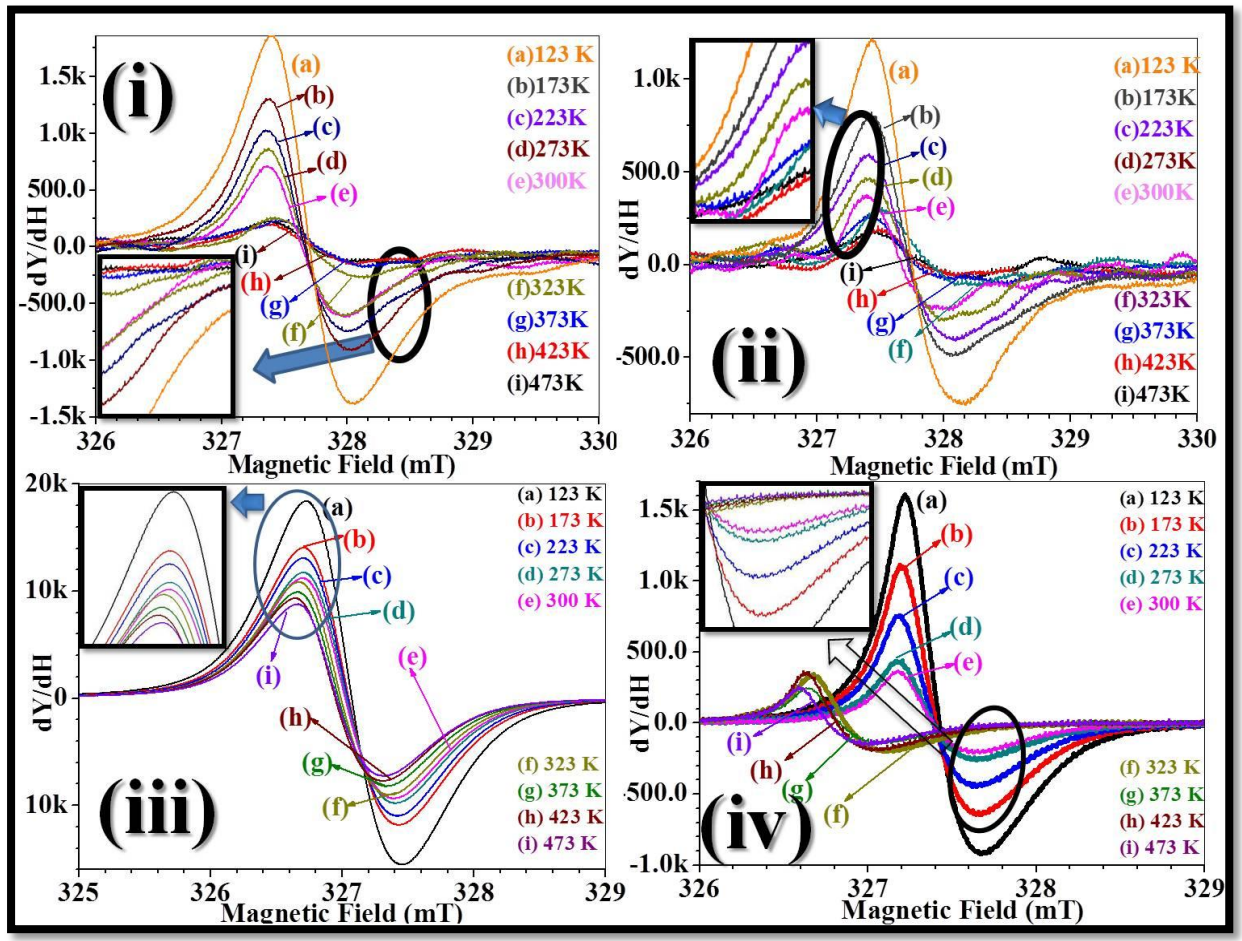

FIG. 1. A comparison of ESR dispersion derivatives, dY/dH, as a function magnetic field for (i) Graphene, (ii) N–Graphene, (iii) GNCs and (iv) N–GNCs over the measured temperature range (a)–(f)123–473 K. For all classes of samples narrowing of linewidth is observed. All samples show unsymmetrical absorption derivative line–shape on low field side, except symmetrical (iii). Open ellipse connected by an arrow shows inset (in each panel) is indicative of hyperfine interactions. The explanation is given in text.

**B. Analysis of spin dynamics of the system**

In the subsequent section, a comprehensive exercise is presented to compute spin transport parameters, using theory of spin relaxation. These parameters may shed light on dynamic of the spin to perform molecular architecting of sp$^2$ network. As a first step towards computing these parameters, peak–to–peak width, $\Delta H_{pp}$, and corresponding g–factor are the crucial parameters. The peak width $\Delta H_{pp}$ has an azimuth angular dependence, θ, i.e. orientation of applied field with respect to the orientation of carbon planes of the samples.

Since, measurements were conducted on the bulk powder specimen then: $\Delta H_{pp}(\theta) = sin^2(\theta)H_{\parallel} + cos^2(\theta)H_{\perp}$. Thus, the measured $\Delta H_{pp}$ consists of contribution from $H_{\parallel}$ as well as $H_{\perp}$. Further, the value of g–factor, at which the resonance has occurred under applied microwave frequency, characterizes the magnetic moment and gyro–magnetic ratio associated with unpaired electrons in the material. If the angular momentum of a system is solely due to spin angular momentum, the tensor g–factor should be isotropic, with the value, 2.00232. Any anisotropy or deviation from this value result involves (i) contributions of the orbital angular momentum from excited state, and (ii) spin–spin contribution. This results into the effective g–factor. Thus, magnitude of effective g–factor is estimated for all the systems under investigation. The effective g–factor is observed to be less than the g–factor for a typically intrinsic spin angular momentum of a free electron i.e. non–degenerated Pauli gas, 2.00232. The difference, $\Delta g$, between 2.00232 and obtained effective g–factor for the system is computed. The observed variation is indicative of corresponding orbital angular magnetic moment could be contributed to the spin magnetic moment. Moreover, the values indicate that, the local magnetic environment in carbon system is distinctly different than that exist in a conventional magnetic material. The computed $\Delta H_{pp}$ and corresponding $\Delta g$ for our systems is enlisted in Table I. In subsequent discussions, temperature averaged magnitude of measured and estimated physical quantities have been quoted using the symbol $<\ldots>_T$. It indicates average of the quantity along with the standard deviation over the measured temperature range. For Graphene, the value of $<\Delta H_{pp}>_T$ is $0.68678 \pm 0.05321$ mT, whereas, for N–Graphene, GNCs, and N–GNCs is, respectively, $0.64131 \pm 0.06356$, $0.70428 \pm 0.02990$, and $0.43470 \pm 0.0268$ (all are in mT). The linewidh of GNCs is observed to be large compared to linewidth of Graphene. After doping nitrogen, there is a linewidth narrowing trend in both the systems, as discussed above. The value of $<\Delta g>_T$ for Graphene is $0.00516 \pm 2.05\times10^{-4}$ and $0.00563 \pm 4.015\times10^{-4}$, for N–Graphene. For GNCs, magnitude of $<\Delta g>_T$ is 0.00455 ± 7.86×10$^{-5}$ and 0.00427 ± 9.48×10$^{-5}$, for N–GNCs. The small values of linewidth, and small deviation in effective g–factor suggest that spin do not originate from transition metal impurities but from only carbon inherited spin species.[26] For Graphene, $\langle\Delta g\rangle_T$ is increased after nitrogen doping, whereas, it is, comparably, smaller for GNCs and reduced after nitrogen doping. This opposite behaviour underlines following facts: (i) orbital angular momentum, and spin–spin contributions are increased in Graphene after doping nitrogen, (ii) the overall strength of these contributions seems to be small in GNCs system and reduced further for N–GNCs, (iii) response of spin to lattice environment, i.e. inversion (Graphene), broken inversion symmetry (GNCs), and with extrinsic adatom, seems to be different, (iv) electrons of nitrogen adatom may also contribute to orbital angular momentum component.

| Temp (in K) | Graphene | | N–Graphene | | GNCs | | N–GNCs | |
|---|---|---|---|---|---|---|---|---|
| | $\nabla H_{pp}$ (in mT) | $\nabla g$ | $\nabla H_{pp}$ (in mT) | $\nabla g$ | $\nabla H_{pp}$ (in mT) | $\nabla g$ | $\nabla H_{pp}$ (in mT) | $\nabla g$ |
| 123 | 0.64390 | 0.00495 | 0.72591 | 0.00518 | 0.73626 | 0.00460 | 0.38851 | 0.00434 |
| 173 | 0.65778 | 0.00506 | 0.60557 | 0.00509 | 0.71285 | 0.00467 | 0.48593 | 0.00425 |
| 223 | 0.69512 | 0.00493 | 0.63612 | 0.00514 | 0.67046 | 0.00463 | 0.43882 | 0.00433 |
| 273 | 0.69522 | 0.00499 | 0.58783 | 0.00562 | 0.71434 | 0.00448 | 0.44144 | 0.00416 |
| 298 | 0.65098 | 0.00578 | 0.59264 | 0.00584 | 0.67123 | 0.00458 | 0.45361 | 0.00439 |
| 323 | 0.69534 | 0.00506 | 0.63001 | 0.00616 | 0.70144 | 0.00445 | 0.43450 | 0.00438 |
| 373 | 0.67963 | 0.00521 | 0.62687 | 0.00600 | 0.70391 | 0.00448 | 0.42721 | 0.00430 |
| 423 | 0.66392 | 0.00535 | 0.62372 | 0.00583 | 0.70637 | 0.00445 | 0.42953 | 0.00421 |
| 473 | 0.81821 | 0.00505 | 0.76981 | 0.00604 | 0.73002 | 0.00457 | 0.41454 | 0.00415 |
| $\langle A\rangle_T$ | 0.68678 ± 0.05321 | 0.00516 ± 2.1×10$^{-4}$ | 0.64131 ± 0.06356 | 0.00563 ± 4.0×10$^{-4}$ | 0.70428 ± 0.0299 | 0.00455 ± 7.86×10$^{-5}$ | 0.43470 ± 0.0268 | 0.00427 ± 9.48×10$^{-5}$ |

Table I. Estimated peak–to–peak linewidth, $\Delta H_{pp}$, and, $\Delta g$, for Graphene and GNCs and their nitrogen doped derivates. The value of $\Delta g$ is difference between g–factor for non–degenerated Pauli gas to g–factor obtained for our samples, at resonance field. The row **$\langle A\rangle_T$** indicate temperature averaged value of $\Delta H_{pp}$ and $\Delta g$.

To evaluate contribution of each component we have carried out analysis on spin–spin relaxation and spin–orbit interactions in the discussions presented below. The magnitude of $\Delta H_{pp}$ bares important information about spin dynamics of the system, specifically, $T_{ss}$, which corresponds to electron spin–spin relaxation time. Due to external perturbation, the deformed spin system regains the state of equilibrium 'up' or 'down' over the characteristic time scale and is termed as spin–spin relaxation time. The entity $\Delta H_{pp}$ and $T_{ss}$ are correlated by equation, $\Delta H_{pp} = \frac{1}{\gamma_e T_{ss}}$, where, $\gamma_e$, has magnitude 1.760859 X $10^{11}$ / sec–T, for electrons. The computed values of $T_{ss}$ are enlisted in Table III. To describe briefly, for Graphene, the values of $T_{ss}$ are varied from 8.82 to 6.94 ps ($<T_{ss}>_T$ = 8.32 ± 0.58 ps), whereas, they are in the range 7.82 to 7.37 ps ($<T_{ss}>_T$ = 8.93 ± 0.80 ps), for N–Graphene. For GNCs, the values are 7.71 to 8.47 ps ($<T_{ss}>_T$ = 8.08 ± 0.265 ps), whereas, for N–GNCs, they are increased to a range 14.62 to 13.77 ps ($<T_{ss}>_T$ = 13.10 ± 0.81 ps). $T_{ss}$ is almost identical for Graphene and GNCs, and after doping nitrogen, there is slight increase for N–Graphene. In contrast, after doping nitrogen in GNCs, $T_{ss}$ is increased to a high value. This indicates that, spin–spin momentum is retarded significantly for N–GNCs.

The principal parameter governing spintronic usability is spin–lattice relaxation time, $T_{sl}$, which quantifies variation of non–thermal spin state around the lattice. The magnitude of $T_{sl}$ is computed using relation: $\frac{1}{T_{sl}} = \frac{28.0\ GHz}{T\ (in\ K)} \times \Delta H_{pp}$. The variation in $T_{sl}$ as a function of temperature for the measured carbon system is shown in Figure 2. For spintronic applications, theoretical estimate for $T_{sl}$ is 1–100 ns, whereas, experimental spin–transport measurements on Graphene showed that $T_{sl}$ is as short as 60–150 psec.[15] FIG. 2 shows, variations in the magnitude of $T_{sl}$ as a function of temperature for the systems under investigation. For Graphene, $T_{sl}$ varies from 6.83 to 22.76 ns, whereas, for N–Graphene observed variation is comparatively marginal from 6.06 to 24.23 ns. The magnitude of $T_{sl}$ is found to varied from 5.97 to 23.14 ns, for GNCs and 11.31 to 41.00 ns, for N–GNCs. For Graphene, N–Graphene and GNCs the values are closer and variations are almost identical. In fact, for these systems, after 300K the value of $T_{sl}$ is almost independent in increase of the temperature. Basically, electron spin relax by transferring energy selectively to those lattice modes with which they resonate. Thus, the resonant modes are on *talking term* with the spins. And one can modify their *cross talk* by introducing break in symmetry inversion (i.e.disorder) or adatom in the $sp^2$ carbon network. However, at higher temperature, the two spins levels seems to be somewhat broadened for the systems other than N–GNCs, resulting a saturation behaviour. In principle, the orbital and the spin angular moment have been considered separately; it is important to know the extent to which these are coupled. As a first approximation the two may be considered independently later introducing a small correction to account for the so called spin–orbit (SO) interaction. For pure, radical free carbon system have, essentially, zero orbital angular momentum; the SO interaction is usually very small for such systems hence for most purposes attention may be focused wholly upon the spin angular momentum. However, SO interaction must necessarily be included in a discussion of the ESR behaviour, as presented below.

SO interaction is one of the most prominent modes of spin relaxation.[11] There are three principal sources of SO coupling, in Graphene: intrinsic, Bychkov–Rashba (BR, related to structural symmetry break) and ripples (related to inevitable wrinkles/folding edges). However, it is not possible to estimate contribution of each component experimentally. Theoretical estimate for intrinsic SOCC ranges 1 μeV–0.2 meV and BR 10–36 μeV/V/nm. Until recently, it has been reported that curved carbon surfaces could have zero–spin splitting SOCC upto 3.4 meV. [R31] The quantification of SO interaction could be done using correlation: $\Delta g = \alpha \left(\frac{L_i}{\Delta}\right)$, where α, is band structure dependent constant ≈ 1, and $L_i$ is spin–orbit coupling constant (SOCC). From the estimated value of $\Delta g$ (in Table I.) and obtaining the magnitude of π–σ bandwidth, $\Delta$, one can estimate $L_i$. A correlation of ESR line–width, $\Delta H_{pp}$, and optical bandwidth, $\Delta$, has already been noted.[27-28] The values of, $\Delta$, for Graphene, N–Graphene, GNCs, and N–GNCs, is respectively, 4.18, 5.27, 4.06, and 3.79 eV. Table II shows variations in the magnitude of $L_i$, over the measured temperature range, for our systems. The magnitude of $L_i$ is found to be three orders of magnitude larger than measured previously.[29]

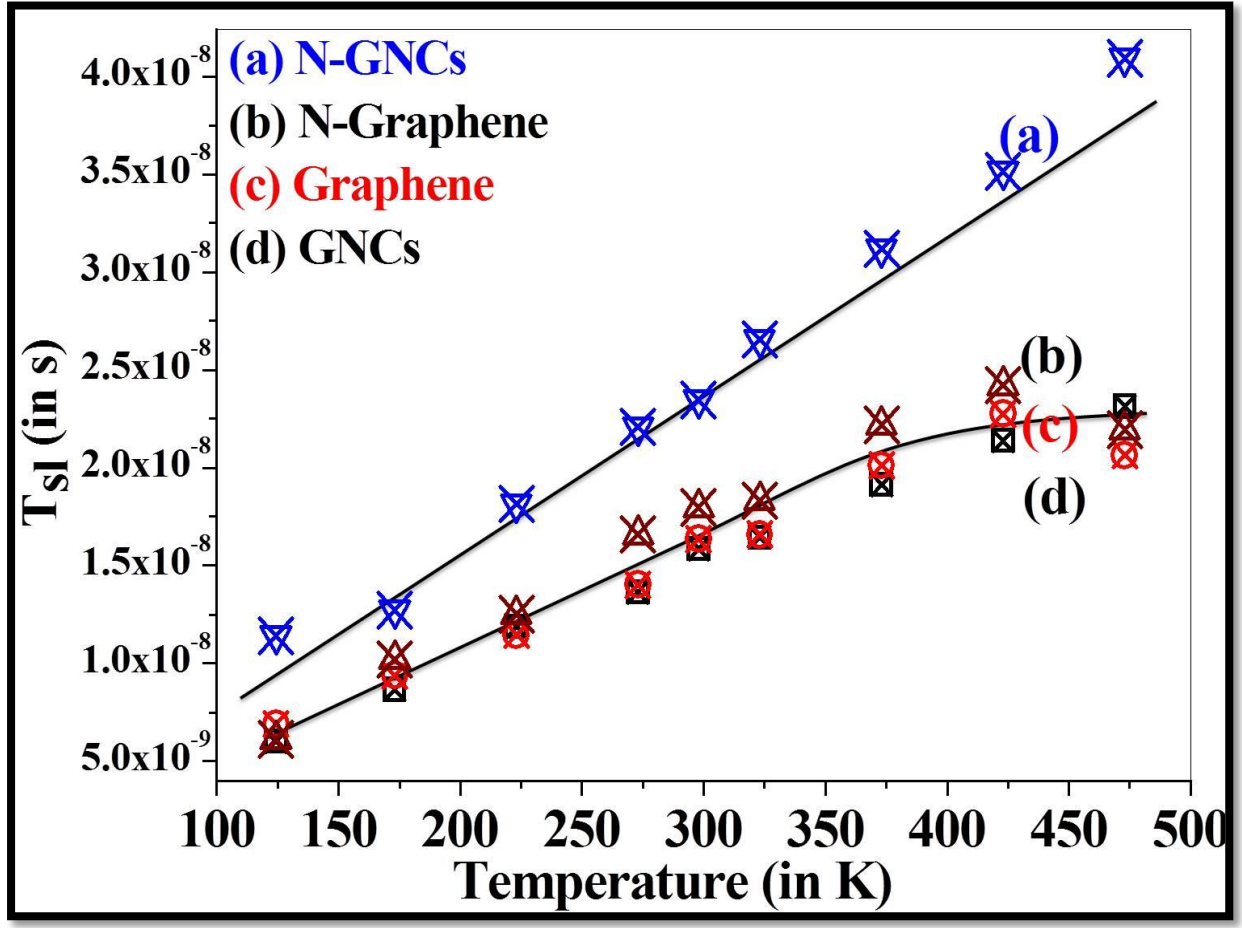

FIG. 2. Variation in spin–lattice relaxation time, $T_{sl}$, over the measured temperature range for (a) N–GNCs, (b) N–Graphene, (c) Graphene, and (d) GNCs. For N–GNCs, almost linear change in $T_{sl}$ is observed from ~ 11 to 41 ns, whereas, for other systems ~ 5 to 25 ns.

| Temp (in K) | Graphene | | N–Graphene | | GNCs | | N–GNCs | |
|---|---|---|---|---|---|---|---|---|
| | SOCC ($L_i$, in meV) | $\Gamma_{spin}$ $\times 10^{-8}$ (eV) | SOCC ($L_i$, in meV) | $\Gamma_{spin}$ $\times 10^{-8}$ (eV) | SOCC ($L_i$, in meV) | $\Gamma_{spin}$ $\times 10^{-8}$ (eV) | SOCC ($L_i$, in meV) | $\Gamma_{spin}$ $\times 10^{-8}$ (eV) |
| 123 | 20.63 | 9.66 | 27.26 | 10.89 | 18.68 | 11.05 | 16.45 | 5.83 |
| 173 | 21.09 | 7.02 | 26.79 | 6.46 | 18.96 | 7.60 | 16.11 | 5.18 |
| 223 | 20.55 | 5.76 | 27.05 | 5.26 | 18.80 | 5.55 | 16.42 | 3.63 |
| 273 | 20.81 | 4.69 | 29.58 | 3.97 | 18.19 | 4.83 | 15.77 | 2.98 |
| 298 | 20.84 | 4.03 | 30.73 | 3.67 | 18.59 | 4.15 | 16.64 | 2.81 |
| 323 | 21.09 | 3.97 | 32.42 | 3.60 | 18.07 | 4.01 | 16.61 | 2.48 |
| 373 | 21.70 | 3.47 | 31.55 | 3.16 | 18.10 | 3.55 | 16.29 | 2.18 |
| 423 | 22.31 | 2.96 | 30.68 | 2.72 | 18.07 | 3.08 | 15.96 | 1.88 |
| 473 | 21.05 | 3.19 | 31.79 | 3.01 | 18.55 | 2.85 | 15.73 | 1.61 |
| $<A>_T$ | 21.271 ± 0.86 | 4.98 ± 2.19 | 29.63 ± 0.21 | 4.75 ± 2.59 | 18.49 ± 0.32 | 5.19 ± 2.63 | 16.17 ± 0.37 | 3.17 ± 1.45 |

TABLE II. Indicating magnitudes of spin–orbit coupling constant, $L_i$ and spin relaxation rate, $\Gamma_{spin}$, computed over the measured temperature range. The last row indicate, **$<A>_T$,** i.e. temperature averaged value of $L_i$ and $\Gamma_{spin}$.

The existent knowledge about the SO interaction in graphene is not yet complete.[4] From the perspectives of solid state qubits construction it is important to understand the role of electronic spin in ordered (Graphene) and disordered (GNCs) $sp^2$ network. Table II shows value of $<L_i>_T$ for Graphene is 21.271 ± 0.86 meV, whereas, it found to be increased to 29.63 ± 0.21 meV for N–Graphene. This indicates that, the orbital angular momentum contribution is increased in Graphene, after nitrogen doping. Comparatively, for GNCs, $<L_i>_T$ is low 18.49 ± 0.32 meV and reduced further to 16.17 ± 0.37 meV, for N–GNCs. The GNCs are disordered $sp^2$ network, in contrast to ordered Graphene. The observed variations could also possibly be attributed to different spin and orbital angular momentum response of nitrogen adatom in ordered and disordered $sp^2$ carbon network leading to different different type of exchange interactions.[4] Thus, analysis of spin–spin ($T_{ss}$) and orbital contribution ($L_i$) together with lattice relaxation ($T_{sl}$) indicates that spin degrees of freedom seems to be somewhat enhanced in N–GNCs. For realistic applications, spin system with enhanced degrees of freedom is advantageous to manipulate, flip and toggle the spin density wave. The SO interaction couples electronic states with opposite spin projections in different bands, typically identified as spin–up and spin–down state and could be quantified as spin–flip parameter, |b|.[30] The perturbation theory set the limit as $|b| \approx L_i/\Delta$. In case of Graphene and allied systems, the SO interaction couples π and σ bands with $L_i \ll \Delta$. FIG. 3 shows variations in spin–flip parameter |b| over the measured temperature range for our systems. The value of |b|, for Graphene, is $4.94 \times 10^{-3}$–$5.04 \times 10^{-3}$ and $5.17 \times 10^{-3}$–$6.03 \times 10^{-3}$, for N–Graphene. For GNCs, it is $4.60 \times 10^{-3}$–$4.57 \times 10^{-3}$ and for N–GNCs $4.35 \times 10^{-3}$–$4.20 \times 10^{-3}$. The values are over the measured temperature range.

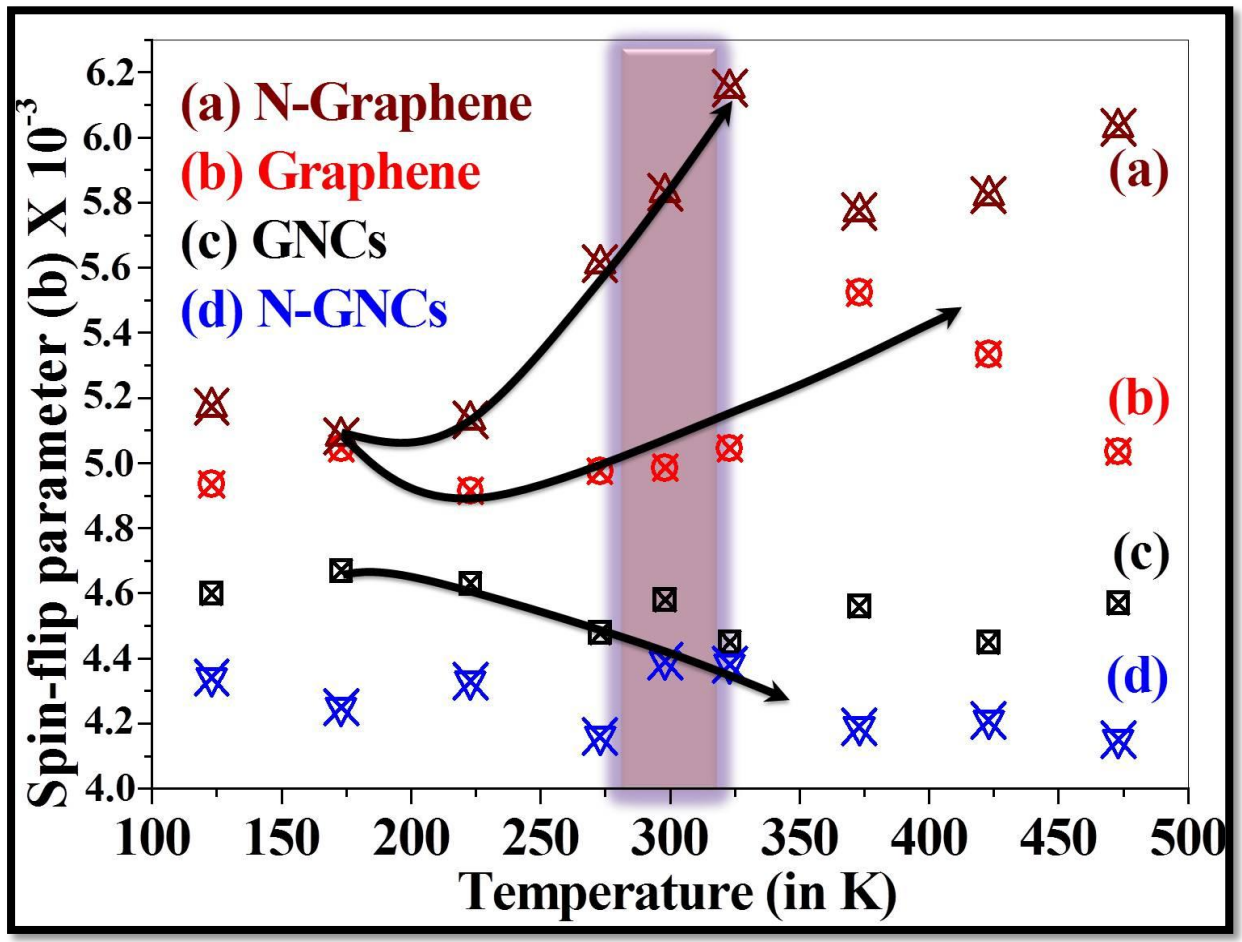

FIG. 3. Spin–flip parameter, |b|, obtained experimentally using Elliot–Yafet relation with eigen Bloch states, $e^{ik \cdot r}$, for spin–Hamiltonian, where, $k$ and $r$ are wave vectors associated with spin density wave. Profile (a) N–Graphene, (b) Graphene, (c) GNCs, and (d) N–GNCs. The shaded region shows room temperature variations in flipping parameter, |b|, for our systems and arrows indicates their opposite behaviour.

The important feature is, |b| has lower value for GNCs (<|b|>$_T$ = 4.55×10$^{-3}$) compared to Graphene (<|b|>$_T$ = 4.55×10$^{-3}$) and reduced further after nitrogen loading in GNCs(<|b|>$_T$ = 4.27×10$^{-3}$). In contrast, <|b|>$_T$ is 5.63×10$^{-3}$ for N–Graphene and value is seems to be increased with respect to <|b|>$_T$ estimated for Graphene. For realistic spin devices, |b| should be conditioned to ~ 25×10$^{-3}$.[32]

Further, the data obtained from electron spin resonance spectra could principally be used to set up spin–Hamiltonian for the system. The spin–Hamiltonian for such system is composed of various terms such as electronic energy (H$_{ele}$), crystal field (H$_{cf,}$), spin–orbit (H$_{SO}$), spin–spin (H$_{SS}$), Zeeman energy(H$_{Ze}$), hyperfine structure(H$_{HFI}$), quadrupole (H$_Q$), and nuclear spin energy, (H$_N$), but, the dominant modes could be H$_{SO}$ and H$_{HFI}$ for the systems under considerations. For the pure carbon systems, H$_{ele}$ and H$_{cf}$ could be neglected due to absence of paramagnetic centres, and radical free environment, whereas, H$_{SS}$ and H$_{Ze}$ could be of the same magnitudes and are orders of magnitudes smaller than H$_{SO}$ and hence neglected. The H$_Q$ and H$_N$ could also be neglected due to the selected temperature region and are predominantly operative only in the sub–mK range. We could add contribution of H$_{HFI}$ due to its presence obtained the ESR line–width, except in GNCs. Thus, the spin–Hamiltonian reads as[33]:

$$[\langle a_k(r)|\uparrow\rangle + \langle b_k(r)|\downarrow\rangle + \prod_i(2I_i+1)^{n_i}]e^{i k \cdot r} \quad (1)$$

$$[\langle a_{-k}(r)^*|\downarrow\rangle + \langle b_{-k}(r)^*|\uparrow\rangle + \prod_i(2I_i+1)^{n_i^*}]e^{i k \cdot r} \quad (2)$$

Where, a and b are complementary sublattice–period of the honeycomb super lattice, and **k**, **r** are density wave vectors, respectively, in real and momentum space. The second term represents nuclear hyperfine interaction at i[th] site and could be neglected for GNCs, but, exist for other three systems.

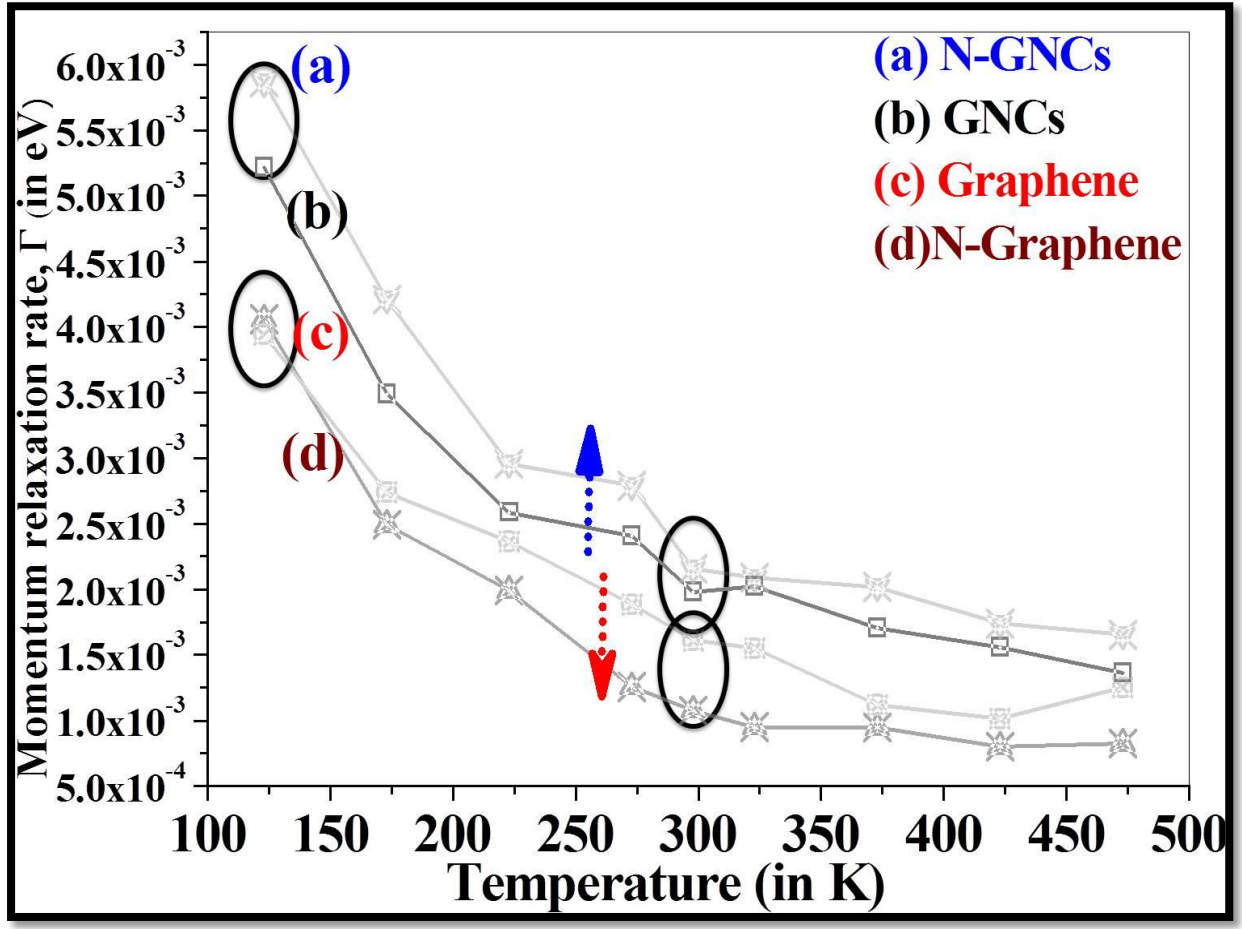

FIG. 4. Estimated momentum relaxation rate, $\Gamma$ (in eV) as a function of temperature (in K) for (a) N–GNCs, (b) GNCs, (c) Graphene, and (d) N–Graphene. Arrows indicate entirely opposite behaviour for Graphene and GNCs. Open ellipses shows room temperature and low temperature behaviour of momentum relaxation rate, $\Gamma$, for the systems.

The spin relaxation rate, $\Gamma_{spin}$, is related to $T_{sl}$ via a relation: $\Gamma_{spin} = \hbar/T_{sl}$, where, $\hbar$ = 6.59 × 10$^{-16}$ eV–s and enlisted in Table II. The relation between, $\Gamma_{spin}$, and momentum relaxation rate, $\Gamma$, is given by : $\Gamma_{spin} = \propto \cdot \left(\frac{L_i}{\Delta}\right)^2 \cdot \Gamma$. FIG. 4 shows variations in $\Gamma$ as a function of temperature for our systems. Our observations for the obtained result are as follows. The value of,Γ, decreases gradually as the temperature increases. For Graphene, Γ, is between 3.94–1.25 meV range, with $<\Gamma>_T$ = 1.94±0.93 meV. For N–Graphene, GNCs and N–GNCs, the range of, Γ, is, respectively, 4.06–0.82 meV ($<\Gamma>_T$ = 1.59±1.09 meV), 5.22–1.36 meV ($<\Gamma>_T$ = 2.25±1.20 meV), and 5.87–1.66 meV ($<\Gamma>_T$ = 2.83±1.39 meV). This indicates that, magnitude of Γ for GNCs is greater than that for Graphene. After doping nitrogen, Γ is increased for GNCs, whereas, it is reduced for N–Graphene. At room temperature, the values of Γ are almost same for GNCs and N–GNCs (indicated by open ellipse). This is indicative of nitrogen loading has no effect on the change in momentum relaxation rate in GNCs system, however, scenario is opposite, for Graphene (open ellipse). At the lowest measured temperature, exact opposite behaviour has been observed (open ellipse). The presence of nitrogen adatom adsorbed and its position in sp$^2$ network has profound effect on the spin configurations of these systems and consequently reflected in the variations in momentum relaxation rate, Γ. The Γ$_{spin}$ has functional dependence on pseudo chemical potential, $\tilde{\mu}$, i.e. local bonding environment and is given by relation (3):

$$\tilde{\mu}(\mu, \Gamma_{\text{spin}}) = \frac{\Gamma_{\text{spin}}}{\pi} \cdot \left(\frac{\mu^2 + \Gamma_{spin}^2}{D^2}\right) + \frac{\Delta}{2}\left(1 - \frac{2}{\pi}\arctan\frac{2\Gamma_{\text{spin}}}{\Delta}\right) \quad (3)$$

Where, μ, is chemical potential and D, continuum cut off parameter is ≈ 3.00 eV. Since $\mu = \Delta/2$, then the magnitude of $\tilde{\mu}$ is estimated for all samples. The pseudo chemical potential, $\tilde{\mu}$, is also connected with the expression of density of states (DOS) ρ ($\tilde{\mu}$, Γ$_{spin}$). The DOS ρ ($\tilde{\mu}$, Γ$_{spin}$) is measured in units of stats/eV–atom and given by:

$$\rho(\tilde{\mu}, \Gamma_{\text{spin}}) = \frac{2 A_c \tilde{\mu}(\mu, \Gamma_{\text{spin}})}{\hbar^2 v_f^2} \quad (4)$$

Where, $A_c = \frac{(5.24 \text{ A}^0)^2}{2}$ per atoms is unit cell for graphene and $v_f$ is Fermi velocity of carriers at Fermi level. Thus, the value of ρ ($\tilde{\mu}$, Γ$_{spin}$) is computed using the estimated values of $\tilde{\mu}$, $\Gamma_{spin}$, as above. It is noteworthy that, for the intrinsic component, when $\tilde{\mu} \gg \Gamma$, $\Gamma(\text{intrinsic}) \approx \frac{L_i^2}{\Delta^2} \Gamma$, which is an Elliot–Yafet–like result.[30] Where, $\Gamma(\text{intrinsic})$ is intrinsic momentum relaxation rate. Further, the ripple relaxation contribution becomes dominant only when $\Gamma \gg \tilde{\mu}$. The comparison of computed values of $\Gamma$ and $\tilde{\mu}$ indicate that, intrinsic, and BR could be the operative coupling channels for our systems. The interdependence of spin dynamic parameters has been studied. FIG. 5 shows (i) variations in $\tilde{\mu}$ with temperature, (ii) DOS variations vs $\Gamma$, (iii) $T_{sl}$ dependence on $\tilde{\mu}$, and (iv) $T_{ss}$ evolution with DOS. The spin orbit coupling allows transition between states of the π–band near the Dirac point with the states from σ–band at the same point and opens–up a gap in the energy dispersion. These transitions imply a change of the spin degree of freedom, i.e. |b|, the spin–flip process resulting into the variations in pseudo chemical potential, $\tilde{\mu}$ of our systems.

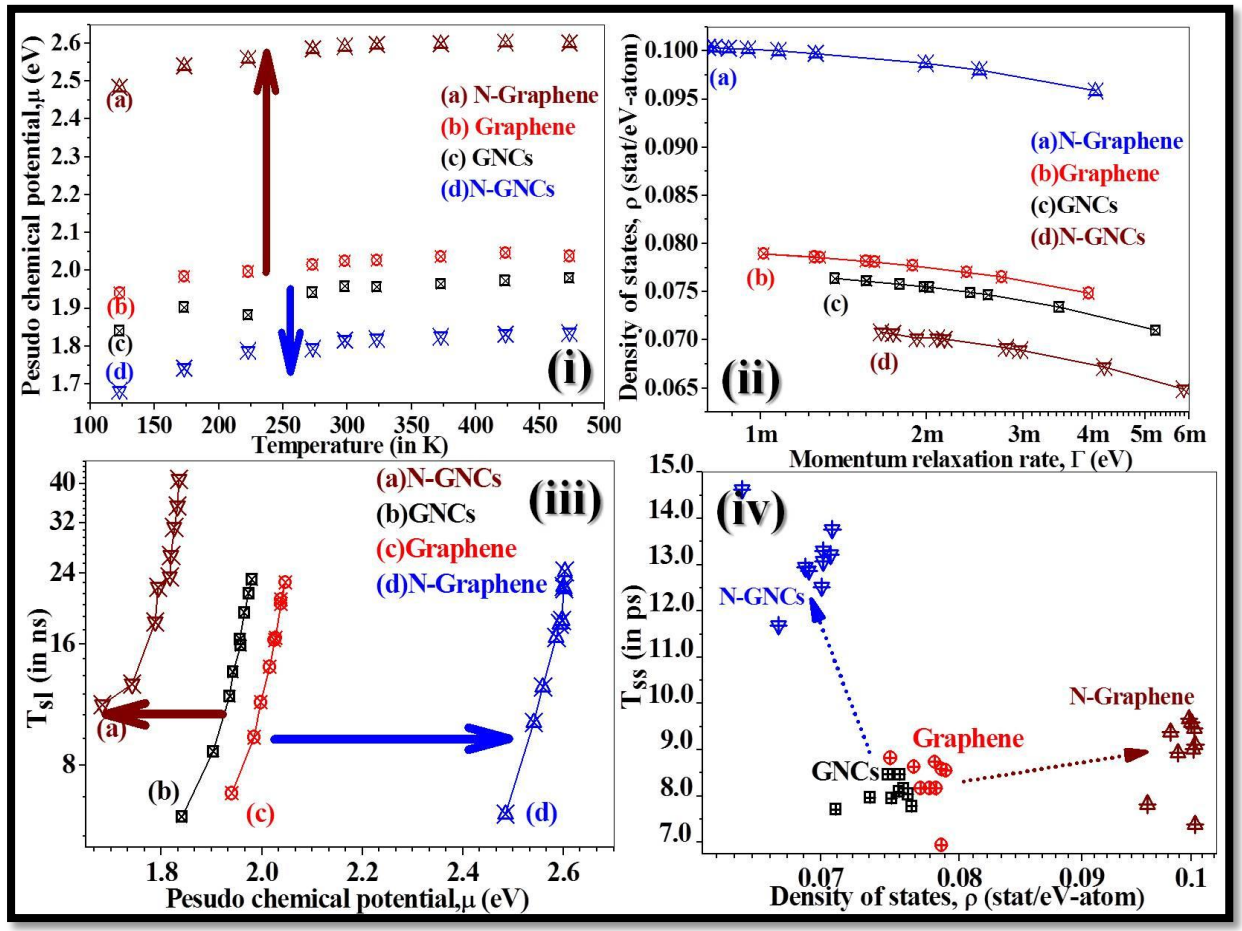

FIG. 5. (i) Pseudo chemical potential, $\tilde{\mu}$, computed for the system, (ii) variations in density of states, $\rho$, as a function of momentum relaxation rate, $\Gamma$, (iii) change in spin-lattice relaxation time, $T_{sl}$ with pseudo chemical potential, $\tilde{\mu}$, and (iv) change in spin-spin relaxation rate, $T_{ss}$, vs density of states, $\rho$. The data is plotted for temperature range 123–473 K. And the arrow indicates contrast behaviour of GNCs and Graphene, after nitrogen loading.

Panel (i) in FIG. 5 shows that, value of $\tilde{\mu}$, for Graphene, ranges 1.9403 to 2.0383 eV with $<\tilde{\mu}>_T$ = 2.0125±0.0337 eV. They are at higher side, for GNCs, which ranges 1.8405 to 1.9804 eV with $<\tilde{\mu}>_T$ = 1.9336±0.04367 eV. This is indicative of midgap opening, due to weaker $<L_i>_T$ for GNCs, compared with the magnitude of $<L_i>_T$ for Graphene. For GNCs, after nitrogen doping, $\tilde{\mu}$ is reduced further and ranges 1.6823 to 1.8349 eV with $<\tilde{\mu}>_T$ = 1.7927±0.0506 eV. This shows the subsequent broadening of midgap, as reflected in further reduction in the obtained $<L_i>_T$ for N–GNCs. In contrast, increase in $\tilde{\mu}$, for N–Graphene, (range : 2.4845 to 2.6015 eV and $<\tilde{\mu}>_T$ = 2.5737±0.03958 eV) indicates the lowering of the midgap with higher value of $<L_i>_T$. The DOS at Fermi level plays crucial role, due the interlink between $\tilde{\mu}$, and $\rho$ via $\Gamma$. From panel (ii) one can see that, the highest DOS are available for N–Graphene compared to other systems. The value of $\rho$ ranges from 0.0958 to 0.10037 stat/eV–atom with mean value $<\rho>_T$ = 0.09929±0.00153 stat/eV–atom. The observed increase in DOS could be due to donation of loan pair of electrons from nitrogen to Graphene. In case of GNCs, each standalone vacancy generates three under coordinated carbon atoms and four dangling electrons keeping $sp^2$ state invariant. This gives rise to four energy levels, three lie near Fermi energy level and fourth lie above the Fermi level. The lowest is fully occupied with two of electrons from dangling bonds keeping third semi filled and upper two fully empty. As a result, one can see the magnitude of DOS is smaller and ranges 0.071 to 0.0764 stat/eV–atom with $<\rho>_T$ = 0.0748 ± 0.00168 stat/eV–atom, for GNCs, in comparison with Graphene (range: 0.07486 to 0.07894 stat/eV–atom and $<\rho>_T$ = 0.0748 ± 0.0013 stat/eV–atom) and its nitrogen derivate. Nitrogen loading further reduces DOS in the range 0.0649 to 0.07049 stat/eV–atom, with mean $\langle\rho\rangle_T$ = 0.06913 ± 0.00193 stat/eV–atom for N–GNCs. The observed decrease could be attributed to exchange interaction that degenerate levels below the Fermi level. Since the spin–polarized levels are derived from the dangling electrons, the spin densities could found to be localized on the undercoordinated atoms. With the assumption of standalone vacancy per three undercoordinated carbon atoms one can found lowest magnitude of effective magnetic moment, $\mu_{eff}$ for GNCs (displayed in Table III.).

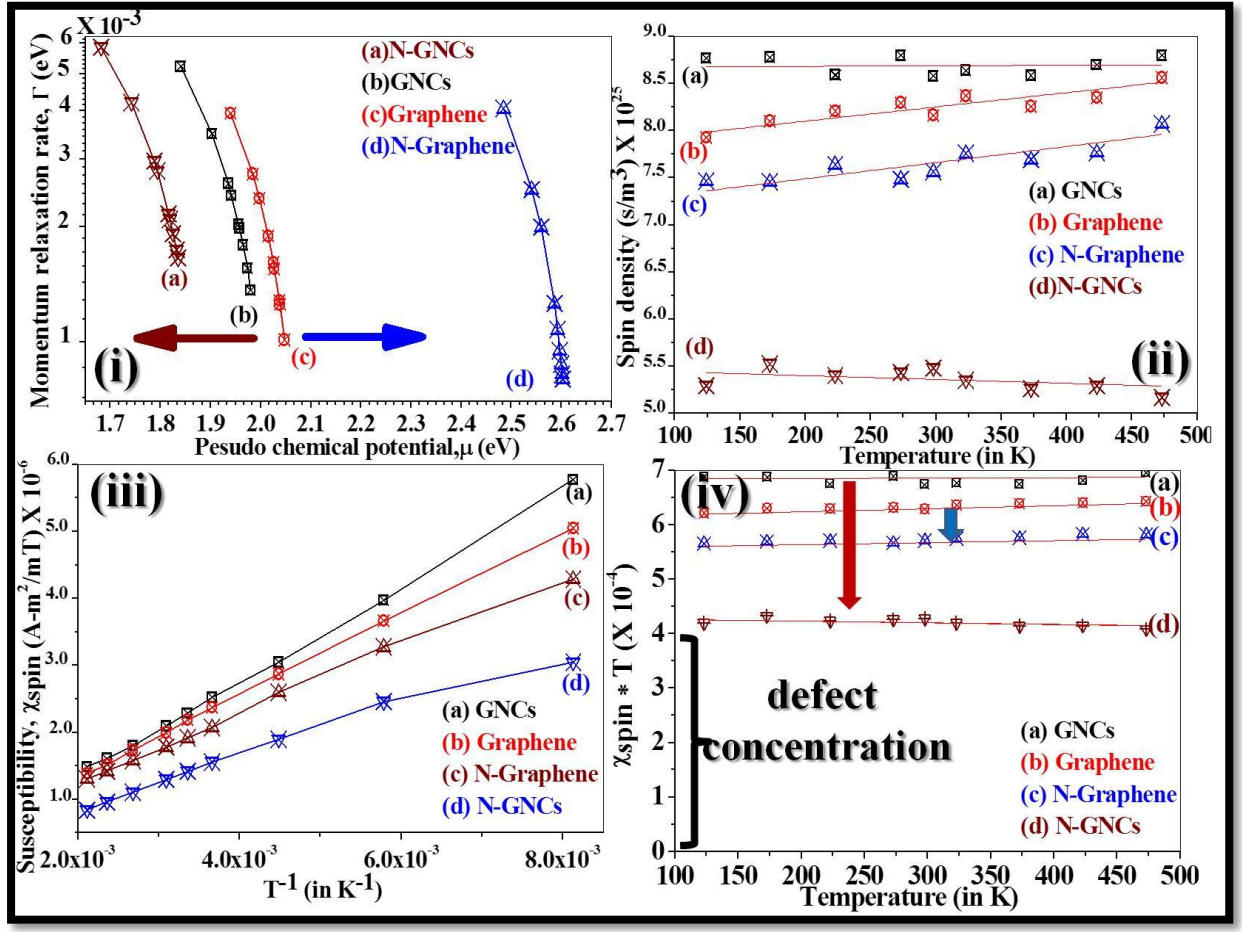

FIG. 6. (i) Change in momentum relaxation rate, $\Gamma$, with pseudo chemical potential, $\tilde{\mu}$, arrows indicate how two systems behaves after nitrogen loading. The data is plotted for the measured temperature range., (ii) computed spin density (s/m$^3$) for Graphene and GNCs and their nitrogen derivatives as a function of temperature, T, (iii) degenerated Pauli spin

susceptibility, $\chi_{spin}$, for spin $\frac{1}{2}$ particles in the system as a function of $T^{-1}$, and (iv) variations in $\chi_{spin}*T$ as a function of T, for the systems, arrows indicate magnitude of change is large for GNCs system compared to Graphene.

Further, lattice–electron, and electron–electron interactions play important role in magnetization of honeycomb carbon.[8] Panel (iii) and (iv) shows variations between $T_{sl}$ vs $\tilde{\mu}$, and $T_{ss}$ vs $\rho$. For N–GNCs, small variations in $\tilde{\mu}$ (1.60 –1.83 eV) caused larger changes in $T_{sl}$ (~ 11–41 ns). The variation seems to be higher compared to other systems. Two prominent features has been observed for the plot of $T_{ss}$ vs $\rho$. First, increase in $\rho$ for Graphene after nitrogen loading with marginal increase in $T_{ss}$, and second increase in $T_{ss}$ with marginal decrease in DOS. However, the observed variations are consistent with discussions made above. FIG. 6 (i) shows variations in $\Gamma$ as a function of $\tilde{\mu}$. The quantity $\Gamma$ has $L_i^2$ dependence. Over the observed variations, the computed curves are for total contribution of $L_i$ i.e intrinsic, BR–type, ripple, and extrinsic (in case of doped systems). However, the extrinsic contribution is added in opposite way for both the system. The ripple relaxation contribution depends on $\Gamma$ only when µ« $\Gamma$, where it resembles an Eilliot−Yafet relation $\Gamma \alpha\ L_{ri}^2\ \Gamma\ \ln\ (D\ /\ \Gamma)$, where D ≈ 3 eV, continuum cut-off parameter. However, in our case $\Gamma$ ranges in meV, whereas, computed $\tilde{\mu}$ is in the range of eV. Thus, one can neglect ripple contribution. Further, the performance of ESR is given by limit of detection of signal i.e. the number concentration of non–degenerated spin $\frac{1}{2}$ particles measured over the temperature. The limit of detection gives a signal–to–noise ratio, S/N, and for our system S/N is 20 for 0.05 mT line–width and the value limit of detection for our system is $5 \times 10^9$ spins/mT. Further, for the ESR system, the gain of the band amplifier and phase–sensitive detector was tuned much smaller than the measured line–width. As a result, the recorded line–shape transformed into the first derivative, $dY/dH$, of the absorption line, Y. By measuring the area, A, under the obtained resonance absorption derivative, $dY/dH$, one can compute the number density of unpaired spins in the samples. The A, usually, is proportional to the spin concentrations (measured in s/m$^3$) in the samples. From FIG. 1, we computed s/m$^3$ for the systems under investigation, using relation[R34]:

$$Y_j = (H_j - H_{j-1})\sum_{i=j}^{m} Y_i' \qquad (5)$$

Where, $Y_j$, $Y_i'$, and $H_j$, is the j$^{th}$ component of absorption line–width, its first derivative, and corresponding field, respectively. FIG. 6(ii) shows variations in spin density as a function of temperature for our systems, moreover, panel (iv) is the complementary data, for defect concentration, obtained from the susceptibility, $\chi_{spin}$ calculations of the samples. One can see that, the concentration of spins for GNCs is varied from 8.77×10$^{25}$ to 8.79×10$^{25}$ with mean $<S_c>_T$ = 8.69×10$^{25}$ ± 9.49×10$^{25}$ s/m$^3$. Whereas, for Graphene, it is in the range 7.93×10$^{25}$–8.57×10$^{25}$ with $<S_c>_T$ = 8.25×10$^{25}$ ± 1.80×10$^{24}$ s/m$^3$. After nitrogen doping, magnitude of spin concentration, $<S_c>_T$, is decreased drastically to 5.36×10$^{25}$ ± 1.15×10$^{24}$ s/m$^3$, for GNCs. However, for N–Graphene the change, $<S_c>_T$, is marginal 7.66 × 10$^{25}$ ± 1.98 × 10$^{24}$ s/m$^3$. Usually, the localized spins are assigned either to unpaired spins of the free radical molecules or defects such as dangling bond, trapped carriers, and vacancies. The exact number could be estimated in the form of Pauli susceptibility of such spins and is given by following relation[30]:

$$\chi_{spin} = \mu_O \, \mu_B^2 \, \frac{N}{V} \cdot (k_B T)^{-1} \qquad (5)$$

Where, $\mu_O$ is permeability of vacuum, $\mu_B$ is Bhor magneton, N/V is spin concentration, k$_B$, Boltzmann's constant, and temperature, T. The variations in $\chi_{spin}$ as a function of T$^{-1}$ is shown in panel (iii) of FIG.6. The temperature dependence of total $\chi_{spin}$ has two components one $\chi_{orb}$ i.e. orbital diamagnetic response and $\chi_n$ due to itinerant nature of electrons and the Pauli paramagnetism due to their spin. At low temperature, $\chi_{spin}$ originates from the DOS. The value of $\chi_{spin}$, for N–GNCs, varies from $3.05 \times 10^{-6}$ A–m$^2$/mT to $8.42 \times 10^{-7}$ A–m$^2$/mT, over the measured temperature range. For N–GNCs, a large deviation has been observed from the linearity, which could be attributed to antiferromagnetic ordering in GNCs, after doping nitrogen. The graphene honeycomb superlattice consists of two complementary sublattices A and B and segregation of conduction electron spin density, n$^c$, over the sublattices generates (i) ferromagnetic ($n_A^c = n_B^c > 0$), (ii) ferrimagnetic ($n_B^c = 0$, $n_A^c > 0$), and (iii) antiferromagnetic ($n_A^c = -n_B^c > 0$) ordering.[R35] The observed behaviour of GNCs indicate that, on each sublattice holes and electrons could get segregates. For, Graphene, $\chi_{spin}$, is 5.05–1.39×10$^{-6}$ A–m$^2$/mT, where as for N–Graphene, and GNCs its value is, respectively, 4.29–1.36×10$^{-6}$ A–m$^2$/mT and 2.74–1.38×10$^{-6}$ A–m$^2$/mT. The vales are quoted over the measured temperature range. Frequently, the Curie law is also plotted in modified form χ∗T as a function of T. Panel (iv) shows variations in χ∗T as a function of T. The extrapolated intersection of each profile in panel (iv) indicates its proportionality to the number of paramagnetic species. For GNCs the value of the intersection is highest, whereas, for N–GNCs it has a lowest magnitude. The effective magnetic moment, μ$_{eff}$, is computed using values of obtained $\chi_{spin}$ for the samples. The Bohr Magneton, μ$_B$, is used as a basic physical constant for magnetic moment of electrons per single atom. The μ$_B$ expresses an intrinsic electron magnetic dipole moment which is ~ 1 Bohr Magneton for electron.[36-37] The study of μ$_B$ of single carbon atom is critical because it verify how many μ$_B$ single carbon atom contains. The effective magnetic moment, $\mu_{eff}$ is computed by use of relation[38]: $\mu_{eff} = 2.83 * [\chi_{spin} * T]^{1/2} * \mu_B$. The computed values are enlisted in Table III.

By and large, our $\mu_{eff}$ calculation shows that, for a cluster of 1000 carbon atoms there are ~ 72 interacting spins for Graphene and after nitrogen doping the value remains almost same ~ 69. For GNCs, this number is somewhat high ~ 74 and reduced substantially to ~ 58 spins for N–GNCs, per thousand carbon atoms. This shows number of electrons participating in spin interaction is reduced significantly, for GNCs, after nitrogen doping. This provides enhanced spin degrees of freedom to spin bath.

| Temp (in K) | Graphene | | N–Graphene | | GNCs | | N–GNCs | |
|---|---|---|---|---|---|---|---|---|
| | $T_{ss}$ (in ps) | $\mu_{eff}$ (in A–m$^2$) | $T_{ss}$ (in ps) | $\mu_{eff}$ (in A–m$^2$) | $T_{ss}$ (in ps) | $\mu_{eff}$ (in A–m$^2$) | $T_{ss}$ (in ps) | $\mu_{eff}$ (in A–m$^2$) |
| 123 | 8.81 | 0.0704 | 7.82 | 0.0661 | 7.71 | 0.075 | 14.60 | 0.055 |
| 173 | 8.63 | 0.0712 | 9.37 | 0.0681 | 7.97 | 0.074 | 11.60 | 0.062 |
| 223 | 8.17 | 0.0692 | 8.92 | 0.0666 | 8.47 | 0.072 | 12.90 | 0.058 |
| 273 | 8.17 | 0.0709 | 9.66 | 0.0673 | 7.95 | 0.074 | 12.80 | 0.058 |
| 298 | 8.73 | 0.0707 | 9.58 | 0.0676 | 8.46 | 0.073 | 12.50 | 0.059 |
| 323 | 8.16 | 0.0732 | 9.01 | 0.0697 | 8.09 | 0.074 | 13.00 | 0.058 |
| 373 | 8.58 | 0.0724 | 9.46 | 0.0696 | 8.17 | 0.073 | 13.30 | 0.057 |
| 423 | 8.55 | 0.0716 | 9.10 | 0.0694 | 8.03 | 0.074 | 13.20 | 0.058 |
| 473 | 6.94 | 0.0794 | 7.37 | 0.0771 | 7.78 | 0.075 | 13.70 | 0.053 |
| <A>$_T$ | 8.32 ± 0.580 | 0.0722 ± 0.0026 | 8.93 ± 0.801 | 0.0691 ± 0.0033 | 8.08 ± 0.265 | 0.0738 ± 9.7×10$^{-5}$ | 13.10 ± 0.819 | 0.0576 ± 0.00251 |

TABLE III: Computed values of spin–spin relaxation time, $T_{ss}$, and effective magnetic moment, $\mu_{eff}$, for Graphene, GNCs and their nitrogen doped derivatives. The last row indicate, <A>$_T$, i.e. temperature averaged value of $T_{ss}$ and $\mu_{eff}$.

Thus, what follows in, from the analyses of spin dynamics? The Graphene and GNCs systems are fundamentally working in opposite way. Another question; which system is well suited for solid state qubit construction?; seems to be N–GNCs. The unit cell of pure Graphene is described by two inequivalent triangular sublattices (A and B) with two independent k–points K and K′. They are having two inequivalent corners of Brillouin zone near the Fermi level which crosses the π–band. This provides an exotic fourfold degeneracy of the low energy spin degenerate states described by two sets of two dimensional chiral spinors which follows Dirac–Wyal equation.[39] It describes the electronic state of the system where K and K′ near the Fermi level is located. Hence, pure Graphene has one electron per carbon atom in the π–band so band below the Fermi level is full (electron like state) and band above the Fermi level is empty (hole like states). Thus, from the perspective of spin degrees of freedom offered to electron, pure Graphene seems to be tight system. But, one could expect more interesting and diversified lattice environment for spin degrees of freedom by GNCs to be offered to electron. Although, GNCs form two dimensional array of six member carbon ring; in contrast to Graphene, these arrays are in the benzenoid carbophane ($sp^2$) connected to $sp^3$ zone via 1,4 hexadine type confirmation.

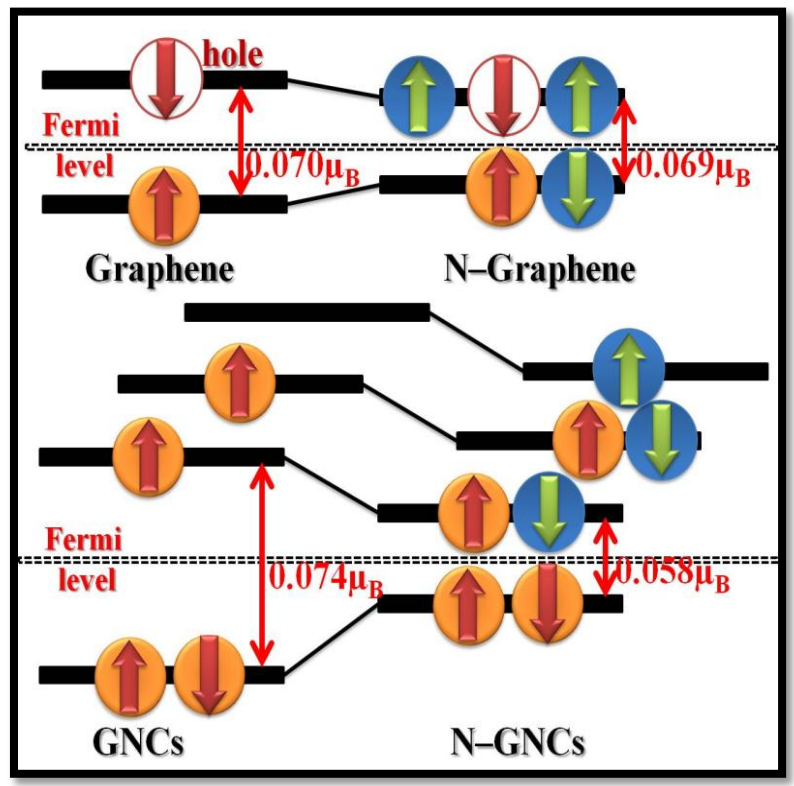

FIG. 7. Schematic depiction of spin splitting around Fermi level of Graphene, GNCs and their nitrogen derivates. For GNCs coupled level split to three levels. The values quoted are effective magnetic moment, $\mu_{eff}$, in terms of $\mu_B$ for their temperature averaged magnitudes.

Thus, GNCs consists of conformational and vacancy disorder. Vacancies also plays important role in disordered system. Once the vacancy is created, the structure is relaxed leaving three under coordinated carbon atoms. These atoms get displaced away from their equilibrium position and move towards their immediate neighbour uniformly along the bonds and each of them will have dangling bonds which are still in the $sp^2$ state. The four dangling bonds gives rise to four energy level three lie near the Fermi energy and fourth lie below the Fermi energy level fully occupied with two of electrons from dangling bond. The remaining three orbital split with the polarized level for spin up and spin down electrons. The remaining two electrons will now occupy the spin up energy levels giving rise to a net magnetic moment.[40] Thus, magnetism, fundamentally, requires an unpaired electron with lattice atom that carries a net, non–zero magnetic moment. These moments can be coupled via exchange interaction with other electrons. The degrees of freedom offered to electron by such system will be more along the interlayer $sp^2+sp^3$ bonded clusters. For N-GNCs, the spin splitting observed to be comparably high. However, the observed decrease in $\mu_{eff}$ in case of N–GNCs shows that the moments, coupled via exchange interactions, are reduced. Electron–electron and lattice–electron interactions play important role in magnetization of honeycomb carbon.[31] [R41][32] Nitrogen donates one of the electrons from its loan pair. The donation is exchange based and exchange takes place with $\frac{1}{2}$ empty $p_z$ orbital. The itinerant–electron gets bound and spin coordinated with localized–π–electron at carbon lattice. The coupling takes place in the screen–columbic repulsive environment. During exchange–coupling, the perturbation offered by uncoordinated itinerant–electron, to empty $p_z$ orbital, could distort σ–bond sensitively. The schematic of spin splitting is shown in FIG. 7. Schematics shows spin exchange coupling of nitrogen electron with hole of carbon system. As a result, the strength of spin–orbit coupling and carrier concentration at Fermi level could be affected. The present analysis indicates that local bond configuration of GNCs after nitrogen doping seems to be favourable for setting up the qubits.

**CONCLUSIONS**

To conclude, what we have studied is the spin dynamics, in Graphene, GNCs, and their nitrogen derivates, over 123 to 473 K, using electron spin resonance technique. Basically, we have four types of carbon lattice environments: (i) system with inversion symmetry (pure $sp^2$ carbon, Graphene), and (ii) broken inversion symmetry and heterostructure (disordered $sp^2$ network, GNCs), (iii) extrinsic nitrogen adatom. The spin dynamics in both the systems are observed to be behaved, somewhat, oppositely after nitrogen doping. Analysis of line width and shape indicated that, for Graphene, line width is found to be varied, asymmetrically, from 1.1027 ± 0.0071 (123K) to 1.2311 ± 0.0063 mT (473K), with increase in the degree of asymmetry for N–Graphene and ranging between 1.1495 ± 0.0066 (123K) to 1.3217 ± 0.0035 mT (473K). The line–widths are found smaller for GNCs (1.2964±0.0083 to 1.2669±0.0091) and narrowed down further for N–GNCs (0.7091±0.0049 to 0.6710±0.0038). The observed variations are attributed to involvement of more than one relaxation time, anisotropic spin orientation of charge inhomogeneities, and hyperfine interactions. For Graphene, $<\Delta H_{pp}>_T$ is 0.68678 ± 0.05321 mT, whereas, for N–Graphene, GNCs, and N–GNCs it is, respectively, 0.64131 ± 0.06356, 0.70428 ± 0.02990, and 0.43470 ± 0.0268 (in mT). The value of $<\Delta g>_T$ for Graphene is 0.00516 ± 2.05×10$^{-4}$ and 0.00563 ± 4.015×10$^{-4}$, for N–Graphene. For GNCs, $<\Delta g>_T$ is 0.00455 ± 7.86×10$^{-5}$ and 0.00427 ± 9.48×10$^{-5}$, for N–GNCs. The small values of linewidths, and analysis of g–factor reveals anisotropic deviation for the systems which indicate variations in the contributions of orbital angular momentum and spin-spin relaxation momentum. The spin-spin relaxation time (Tss), is $<T_{ss}>_T$ is 8.32 ± 0.58 ps, for Graphene, whereas, for N–Graphene, 8.93 ± 0.80 ps, for GNCs, 8.08 ± 0.265 ps, and increased, for N–GNCs, upto 13.10 ± 0.81 ps. The spin-lattice relaxation time ($T_{sl}$), for Graphene, is varied from 6.83 to 22.76 ns, whereas, for N–Graphene from 6.06 to 24.23 ns, for GNCs, 5.97 to 23.14 ns and 11.31 to 41.00 ns, for N–GNCs. The extent of orbital and spin angular moment coupling have been studied by estimating spin–orbit coupling constant ($L_i$). The estimated $<L_i>_T$ is 18.49 ± 0.32 meV for GNCs and reduced further, to 16.17 ± 0.37 meV, after nitrogen doping; however, in contrast, for Graphene its value is high (21.271 ± 0.86 meV) and increased upto 29.63 ± 0.21 meV, for N–Graphene. This is primary indication that, N-GNCs is a system with enhanced spin degrees of freedom. The computed value of spin flip parameter (|b|) is $4.94 \times 10^{-3}$–$5.04 \times 10^{-3}$, for Graphene, and $5.17 \times 10^{-3}$–$6.03 \times 10^{-3}$, for N–Graphene. For GNCs, it is in low range $4.60 \times 10^{-3}$–$4.57 \times 10^{-3}$ and from realistic view point, for N–GNCs, $4.35 \times 10^{-3}$–$4.20 \times 10^{-3}$. The spin relaxation rate ($\Gamma_{spin}$) is observed to be decreased gradually over 123 to 473 K. For Graphene, $<\Gamma_{spin}>_T$ is $4.98 \pm 2.19 \times 10^{-8}$ eV, for N-Graphene, $4.75 \pm 2.59 \times 10^{-8}$ eV, for GNCs, $5.19 \pm 2.63 \times 10^{-8}$ eV, for N–GNCs, $3.17 \pm 1.45 \times 10^{-8}$ eV. This indicates that, spin rates are quiet retarded for N–GNCs. The momentum relaxation rate, $<\Gamma>_T$ is $1.94 \pm 0.93$ meV, for Graphene, whereas, for N–Graphene, GNCs and N–GNCs, respectively, $1.59 \pm 1.09$ meV, $2.25 \pm 1.20$ meV, and $2.83 \pm 1.39$ meV. The $\Gamma_{spin}$ has functional dependence on pseudo chemical potential ($\tilde{\mu}$) with the condition $\tilde{\mu} \gg \Gamma$. The computed value of $<\tilde{\mu}>_T$ for Graphene is $2.0125 \pm 0.0337$ eV, $1.9336 \pm 0.04367$ eV for GNCs, $1.7927 \pm 0.0506$ eV, for N–GNCs and $2.5737 \pm 0.03958$ eV, for N–Graphene. The observed changes are attributed to variations in the midgap states. This has effect on density of states, the value of $<\rho>_T$ is $0.0748 \pm 0.00168$ states/eV–atom for GNCs, however, for N–GNCs, it is reduced to $0.06913 \pm 0.00193$ states/eV–atom. For Graphene, it is $0.07764 \pm 0.0013$ states/eV–atom and enhanced to $0.09929 \pm 0.00193$ states/eV–atom for N–Graphene. The spin concentration $<S_c>_T$ for Graphene is found to be lower $8.25 \times 10^{25} \pm 1.80 \times 10^{24}$ s/m³, with ~ 70 interacting spins per thousand carbon atoms and reduced marginally to $7.66 \times 10^{25} \pm 1.98 \times 10^{24}$ s/m³, with almost same number of interacting spins. For GNCs, the variations are drastic, it indicate that $<S_c>_T$ is $8.69\times10^{25} \pm 9.49\times10^{25}$ s/m$^3$ with ~ 74 interacting spins per thousand carbon atoms, and with nitrogen doping it reduced to $5.36\times10^{25} \pm 1.15\times10^{24}$ s/m$^3$ with ~ 57 interacting spins per same number of carbon atoms. The spin susceptibility, for N–GNCs, is deviated largely from linearity which leads us to a conclusion that antiferromagnetism could be the operative channel in N–GNCs. . For, Graphene, $\chi_{spin}$, is 5.05–1.39×10$^{-6}$ A–m$^2$/mT, where as for N–Graphene, and GNCs its value is, respectively, 4.29–1.36×10$^{-6}$ A–m$^2$/mT and 2.74–1.38×10$^{-6}$ A–m$^2$/mT. For realistic qubit design, transport parameters such as spin–orbit coupling constant, spin–lattice relaxation time, and spin–flip process should be optimum with optimum spin dynamics, at room temperature. Analysis gives us the clue to design qubit and it seems that N–GNCs could be the well suited candidate for this.

## ACKNOWLEDGMENT

AK and PSA are thankful to the Ministry of Defence, Govt. of India for giving grant through ER–IPR, DIAT–DRDO nanoproject.